\documentstyle[aps,multicol,epsf]{revtex}

\begin{document}

\draft

\title{Shell model studies of the proton drip line nucleus $^{106}$Sb}

\author{T.\ Engeland, M.\ Hjorth-Jensen and E.\ Osnes}

\address{Department of Physics,
         University of Oslo, N-0316 Oslo, Norway}

\maketitle

\begin{abstract}

We present results of shell model calculations for the proton drip
line nucleus $^{106}$Sb.
The shell model calculations were performed based on an effective
interaction for the $2s1d0g_{7/2}0h11_{11/2}$ shells employing 
modern models for the nucleon-nucleon interaction. 
The results are compared with the recently 
proposed experimental yrast states. 
A good agreement with experiment 
is found lending support to the experimental
spin assignements. 
\end{abstract}

\pacs{PACS number(s): 21.60.-n, 21.60.Cs, 27.60.+j}

\begin{multicols}{2}
Considerable attention is at present being devoted to the
experimental and theoretical study of nuclei close to the limits
of stability. Recently, heavy neutron deficient nuclei in the mass
regions of $A=100$ have been studied, and nuclei like $^{100}$Sn
and neighboring isotopes have been identified \cite{sn100a,sn100b,sn100c}.
Moreover, the proton drip line has been established in the $A=100$
and $A=150$ regions \cite{protondripline} and nuclei like
$^{105}$Sb and $^{109}$I have recently  been established as  ground-state 
proton emitters \cite{sb105,i109}.
The next to drip line nucleus for the antimony isotopes, $^{106}$Sb
with a proton separation energy of $\sim 400$ keV, was studied recently
in two experiments and a level scheme for the yrast states was proposed
in Ref.~\cite{sb106}.

The aim of this work is thus to see whether shell-model calculations,
which
employ realistic effective interactions based on state of the art 
models for the nucleon-nucleon interaction,  
are capable of reproducing
the experimental results for systems close to the stability line.
Before we present  our results, we will briefly 
review  our theoretical framework.
In addition, we present results
for effective proton and neutron charges based on perturbative
many-body methods. These effective charges will in turn
be used in a shell-model analysis of $E2$ transitions.

The aim of microscopic nuclear structure calculations is to derive
various properties of finite nuclei from the underlying 
hamiltonian describing the interaction between 
nucleons. 
We derive an appropriate 
effective two-body interaction for valence neutrons and protons  in the 
single-particle orbits $2s_{1/2}$, $1d_{5/2}$, $1d_{3/2}$, 
$0g_{7/2}$ and $0h_{11/2}$. As closed shell core we use $^{100}$Sn.
This effective two-particle interaction is in turn used in the 
shell model model study of $^{106}$Sb.
The shell model problem requires the solution of a real symmetric
$n \times n$ matrix eigenvalue equation
\begin{equation}
       \tilde{H}\left | \Psi_k\right\rangle  = 
       E_k \left | \Psi_k\right\rangle ,
       \label{eq:shell_model}
\end{equation}
with $k = 1,\ldots, K$. 
At present our basic approach to 
finding solutions to Eq.\ (\ref{eq:shell_model})
is the Lanczos algorithm, an iterative method which gives the solution of
the lowest eigenstates. The 
technique is described in detail in Ref.\ \cite{whit77}, 
see also Ref.\ \cite{ehho95}. 

To derive the effective interaction, we employ 
a  perturbative many-body scheme starting with the free nucleon-nucleon
interaction. This interaction is in turn renormalized
taking into account the specific nuclear medium. The medium
renormalized potential, the so-called $G$-matrix, is then
employed in a perturbative many-body scheme, as detailed in
Ref.\ \cite{hko95} and reviewed briefly below. 
The bare nucleon-nucleon interaction we use is the charge-dependent
meson-exchange model of Machleidt and co-workers \cite{cdbonn}, the so-called
CD-Bonn model.
The potential model of Ref.\ \cite{cdbonn} is an extension of the 
one-boson-exchange models of the Bonn group \cite{mac89}, where mesons 
like $\pi$, $\rho$, $\eta$, $\delta$, $\omega$ and the fictitious
$\sigma$ meson are included. In the charge-dependent version
of Ref.\ \cite{cdbonn}, the first five mesons have the same set
of parameters for all partial waves, whereas the parameters of
the $\sigma$ meson are allowed to vary. 

The first step 
in our perturbative many-body scheme is to handle 
the fact that the repulsive core of the nucleon-nucleon potential $V$
is unsuitable for perturbative approaches. This problem is overcome
by introducing the reaction matrix $G$ given by the solution of the
Bethe-Goldstone equation
\begin{equation}
    G=V+V\frac{Q}{\omega - QTQ}G,
\end{equation}
where $\omega$ is the unperturbed energy of the interacting nucleons,
and $H_0$ is the unperturbed hamiltonian. 
The operator $Q$, commonly referred to
as the Pauli operator, is a projection operator which prevents the
interacting nucleons from scattering into states occupied by other nucleons.
In this work we solve the Bethe-Goldstone equation for five starting
energies $\omega$, by way of the so-called double-partitioning scheme
discussed in e.g.,  Ref.\ \cite{hko95}. 
A harmonic-oscillator basis was chosen for the
single-particle
wave functions, with an oscillator energy $\hbar\Omega$ given
by
$\hbar\Omega = 45A^{-1/3} - 25A^{-2/3}=8.5 $ MeV,  $A=100$ being the mass
number.

Finally, we briefly sketch how to calculate an effective 
two-body interaction for the chosen model space
in terms of the $G$-matrix.  Since the $G$-matrix represents just
the summmation to all orders of ladder diagrams with particle-particle
intermediate states, there are obviously other terms which need to be included
in an effective interaction. Long-range effects represented by 
core-polarization terms are also needed.
The first step then is to define the so-called $\hat{Q}$-box given by
\begin{eqnarray}
   &P\hat{Q}P=PGP +\nonumber\\
   &P\left(G\frac{Q}{\omega-H_{0}}G+ G
   \frac{Q}{\omega-H_{0}}G \frac{Q}{\omega-H_{0}}G +\dots\right)P.
   \label{eq:qbox}
\end{eqnarray}
The $\hat{Q}$-box is made up of non-folded diagrams which are irreducible
and valence linked. The projection operators $P$ and $Q$ define the model
space and the excluded space, respectively, with $P+Q=I$. 
All non-folded diagrams through 
third order in the interaction $G$ are included in the definition 
of the $\hat{Q}$-box while so-called 
folded diagrams are 
included to infinite order through the summation scheme discussed
in Refs.\ \cite{hko95,ko90}.

Effective interactions based on 
the CD-Bonn nucleon-nucleon interaction 
have been used by us for several mass regions, and give in general
a very good agreement with the data, see Refs.\ 
\cite{alex98,ehho98,ssh99,david99}.

In addition to deriving an effective interaction for the shell
model, we present also effective proton and neutron charges 
based on our perturbative many-body methods. These charges
are used in our studies of available $E2$ data below.
In this way, degrees
of freedom not accounted for by the shell-model space are partly
included through the introduction of an effective charge.
The effective single-particle operators for the effective charge 
are calculated along the same lines  
as the effective interaction. In   
nuclear transitions, the quantity of 
interest is the transition matrix element between an initial state 
$\left|\Psi_i\right\rangle$ and a final state $\left|\Psi_f\right\rangle$ 
of an operator ${\cal O}$ (here it is the $E2$ operator) defined as 
\begin{equation} 
               {\cal O}_{fi}= 
               \frac{\left\langle\Psi_f\right| 
               {\cal O}\left|\Psi_i\right\rangle } 
               {\sqrt{\left\langle\Psi_f | \Psi_f \right\rangle 
               \left\langle \Psi_i | \Psi_i \right\rangle}}. 
               \label{eq:effop1} 
\end{equation} 
Since we perform our calculation in a reduced space, the exact 
wave functions $\Psi_{f,i}$ are not known, only their 
projections $\Phi_{f,i}$ onto the model space. We are then confronted with the 
problem of evaluating ${\cal O}_{fi}$ when only the model 
space wave functions are known. In treating this problem, it is usual 
to introduce an effective operator 
${\cal O}_{\mathrm{eff}}$ different from the original operator ${\cal O}$ 
defined by requiring 
\begin{equation} 
           {\cal O}_{fi}=\left\langle\Phi_f\right |{\cal O}_{\mathrm{eff}} 
           \left|\Phi_i\right\rangle. 
\end{equation} 
The standard 
scheme is then to employ a  
perturbative expansion for the effective operator, see e.g.\ Refs.\  
\cite{towner87,eo77,kren77}. 
 
To obtain effective charges, we 
evaluate all effective operator diagrams through second-order,
excluding Hartree-Fock insertions,  in the 
$G$-matrix obtained with  the CD-Bonn interaction. Such diagrams 
are discussed in the reviews by Towner \cite{towner87} 
and Ellis and Osnes \cite{eo77}.   
The state dependent effective charges are listed in Table
\ref{tab:charges} for the diagonal contributions only.
In order to reproduce the experimental $B(E2;4_1^+\rightarrow 2_1^+)$
transition of Ref.\ \cite{sb106}, the authors introduced 
effective charges $e_p=1.72e$ and
$e_n=1.44e$ for protons and neutrons, respectively.
We see from Table \ref{tab:charges} that the microscopically calculated
values differ significantly from the above values
from  Ref.\ \cite{sb106}. This
could however very well be an artefact of the chosen model space and 
effective interaction employed in the shell model analysis of
Ref.\ \cite{sb106}. The reader should also keep in mind that our
model for the single-particle wave functions, namely the 
harmonic oscillator, may not be the most appropriate for the proton
single-particle states, 
since the proton separation energy is of the order of some
few keV. 
When compared with the theoretical calculation of Sagawa {\em et al.} \cite{ssbw87},
our neutron effective charges agree well with theirs, whereas the proton effective charge
deduced in Ref.\ \cite{ssbw87} is slightly larger, $e_p\sim 1.4e$. We note also that in the 
Hartree-Fock calculation with a Skyrme interaction and 
accounting for effects from the continuum, 
Hamamoto and Sagawa \cite{hs97} obtained
effective charges of $e_n=1.35e$  and $e_p=1.0e$ for $^{100}$Sn. 
Below we will allow the effective charges
to vary in order to  reproduce as far as possible
the  experimental value of $2.8(3)$ W.u.\  for the transition 
$B(E2;4_1^+\rightarrow 2_1^+)$. There we will also relate the theoretical
values for the effective charges to those extracted from data
around $A=100$, see e.g., Ref.\ \cite{matej}.

The calculations were performed with two possible 
model spaces, one which comprises all single-particle orbitals of the 
$1d_{5/2}0g_{7/2}1d_{3/2}2s_{1/2}0h_{11/2}$ shell and one which excludes
the $0h_{11/2}$ orbit. 
The latter model space was employed by the authors of Ref.\ \cite{sb106}
in their shell model studies.
Since the single-neutron and single-proton
energies with respect to $^{100}$Sn are not well-established, we have adopted
for neutrons the same single-particle energies as used in large-scale
shell-model calculations of the Sn isotopes, see Refs.\ \cite{ehho98}.
The neutron single-particle energies are
$\varepsilon_{0g_{7/2}}-\varepsilon_{1d_{5/2}}=0.2$ MeV,
$\varepsilon_{1d_{3/2}}-\varepsilon_{1d_{5/2}}=2.55$ MeV,
$\varepsilon_{2s_{1/2}}-\varepsilon_{1d_{5/2}}=2.45$ MeV and
$\varepsilon_{0h_{11/2}}-\varepsilon_{1d_{5/2}}=3.2$ MeV.
These energies, when employed with our effective interaction described
above, gave excellent results for both even and odd
tin isotopes from $^{102}$Sn to $^{116}$Sn. 
The proton single-particle energies are less established and we simply adopt 
those for the neutrons. Since the proton separation energy is of the order
of $\sim 400$ kev, it should suffice to carry out a shell-model
calculation with just the $1d_{5/2}0g_{7/2}$ orbits for protons.
The total wave function, see the discussion below, is however
to a large extent 
dominated by the $1d_{5/2}$ orbital for protons, with small admixtures
from the $0g_{7/2}$ proton orbital. The influence from the other 
proton orbits is thus minimal. 

\begin{figure}
   \setlength{\unitlength}{1mm}
   \begin{picture}(140,180)
   \put(-20,0){\epsfxsize=18cm \epsfbox{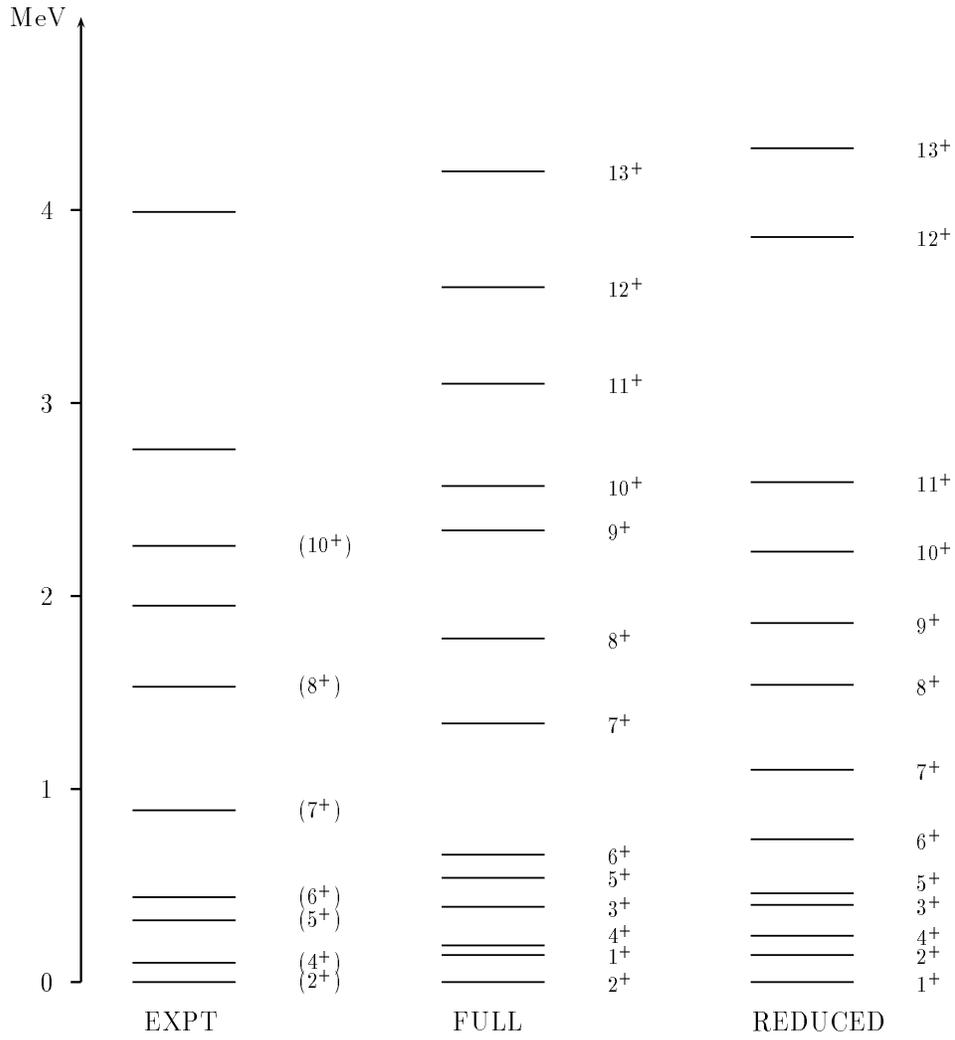}}
   \end{picture}
\caption{Low-lying states for  $^{106}$Sb, theory and experiment. 
         Energies in MeV. FULL means the model space which comprises all
         single-particle orbits, $2s1d0g_{7/2}0h11_{11/2}$. REDUCED means
         that the $0h11_{11/2}$ single-particle orbit is not included.}
\label{fig:energies}
\end{figure}

The resulting eigenvalues are displayed in Fig.\ \ref{fig:energies}
for the two choices of model space together with the experimental
levels reported in Ref.\ \cite{sb106}. Not all experimental
levels have been given a spin assignment and all experimental spin values
are tentative.  
The label FULL stands for the model space which includes all orbits
from the $1d_{5/2}0g_{7/2}1d_{3/2}2s_{1/2}0h_{11/2}$ shell while
REDUCED stands for the model space where the $0h_{11/2}$ orbit has been
omitted.
As can be seen from Fig.\ \ref{fig:energies}, the agreement with
experiment is also rather good, with the model space which includes
all orbitals of the $1d_{5/2}0g_{7/2}1d_{3/2}2s_{1/2}0h_{11/2}$ shells
being closest to the experimental level assignements.
The reader should however note that in Ref.\ \cite{sb106} it is not
excluded that the ground state could have spin $1^+$, which means
that the experimental spin values in Fig.\ \ref{fig:energies} 
should be reduced by 1. In our theoretical calculations we obtain in
addition to a state with spin $1^+$, also a state with $3^+$ not seen
in the experiment of Ref.\ \cite{sb106}. In case the ground state
turns out to have spin $1^+$, the reduced model space in our calculations
would yield a better agreement with the data.

\begin{figure}
   \setlength{\unitlength}{1mm}
   \begin{picture}(140,180)
   \put(-20,0){\epsfxsize=18cm \epsfbox{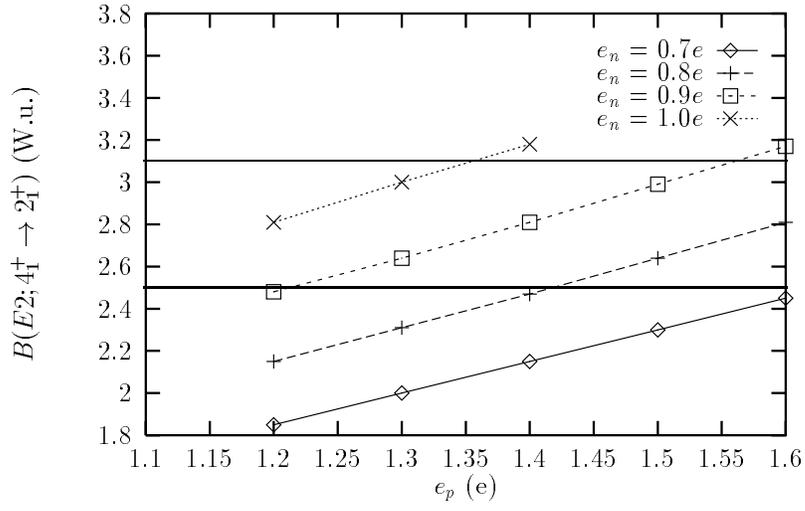}}
   \end{picture}
\caption{Value for the $B(E2;4_1^+\rightarrow 2_1^+)$ transition as function of
         different effective charges in units of W.u. The horizontal
         lines represent the experimental window, with a value of 
         $2.8 \pm 0.3$ W.u.}
\label{fig:empe2}
\end{figure}

The wave functions for the various states are to a large extent
dominated by the $0g_{7/2}$ and $1d_{5/2}$ single-particle orbits
for neutrons ($\nu$)
and the $1d_{5/2}$ single-particle
orbit for protons ($\pi$). 
The $\nu 0g_{7/2}$ and $\nu 1d_{5/2}$ single-particle orbits represent
in general more than $\sim 90\%$ of the total neutron single-particle
occupancy, while the  $\pi 1d_{5/2}$ single-particle orbits stands for 
$\sim 80-90\%$ of the proton single-particle occupancy. The other single-particle
orbits play  an almost negligible role in the structure of the wave functions.
The only notable exception is the $7_1^{+}$ state where $\pi 0g_{7/2}$
stands for the $84\%$ of the proton single-particle occupancy.
This has also important repercussions on the contributions to the
measured $E2$ transition $B(E2;4_1^+\rightarrow 2_1^+)$, where  the stucture of the 
wave functions of the  
$4_1^+$ and $2_1^+$ states are to  a large extent dominated by the 
$\nu 1d_{5/2}$ and $\pi 1d_{5/2}$ single-particle orbits. The $0g_{7/2}$ orbits
play a less significant  role in the structure of the wave functions, 
and since the $0g_{7/2}\leftrightarrow 1d_{5/2}$ transition matrix element
tends to be weaker
than the  one  between $1d_{5/2}\leftrightarrow 1d_{5/2}$, the $E2$
transition will be dominated by  the latter contributions.
As also noted by Sohler {\em et al.} \cite{sb106}, the $E2$ transition is dominated by
neutron contributions.  This can also be seen from Fig.\  \ref{fig:empe2} 
where we show the result for the above $E2$ transition as function of 
different choices for the effective charges. 
We see that the largest change in the value of the $E2$ transition takes place
when we vary the effective charge of the neutron, whereas when the proton charge is
changed, the percentual change is smaller. 
Furthermore, if we use the largest values for effective charges of Table
\ref{tab:charges}, namely $e_n=0.72e$ and $e_p=1.16e$, we obtain $1.84$ W.u.\ for the
$E2$ transition. Compared with the experimental value
of $2.8(3)$ W.u.\ this may indicate that both the proton and the neutron effective
charges should be slightly increased. From Fig.\ 
\ref{fig:empe2} we see that effective charges of $e_n=0.9e $ and 
$e_p=1.4e\pm 0.2e$
seem to yield the best agreement with experiment, although
neutron charges of $e_n=0.8e$ and $e_n=1.0e$ yield results
within the experimental window of Fig.\ \ref{fig:empe2} . 
The neutron effective charges
would agree partly with those exctracted from the data in the Sn isotopes 
\cite{ssbw87} and from theoretical
calculations of $E2$ transitions in heavy Sn isotopes \cite{ehho98}, where a value
$e_n\sim 1$ is adopted in order to reproduce the data.  
The calculated effective charges of 
Table \ref{tab:charges} are however on the lower side.
However, the deduced effective charges from the 
$B(E2;6_1^+\rightarrow 4_1^+)$ transitions in $^{102}$Sn \cite{matej} and $^{104}$Sn
\cite{hubert95} indicate that $e_n\sim 1.6-2.3e$, depending
on the effective interaction employed in the shell-model analyses. Clearly, the effective
interaction which is used, and its pertinent model space, approximations
made in the many-body formalism etc., will influence the extraction of effective charges.
This notable difference in the effective charges could be due to the fact that
the  $B(E2;6_1^+\rightarrow 4_1^+)$ transitions in $^{102}$Sn and $^{104}$Sn
involve configurations not accounted for by the 
$1d_{5/2}0g_{7/2}1d_{3/2}2s_{1/2}0h_{11/2}$ model space.
A proton effective charge of $e_p\sim 1.4e$ is close to values inferred
from experiment for the $N=50$ isotones, see Ref.\ \cite{sn100a,sn100b}, shell-model
calculations of $E2$ transitions for the $N=82$ isotones \cite{hehos97} and
the theoretical estimates of Ref.\ \cite{ssbw87}.

In summary, a shell-model calculation with realistic effective
interactions of the newly reported low-lying
yrast states of  the proton drip line nucleus $^{106}$Sb, reproduces well
the experimental data. Since the wave functions of the various states are to a large
extent dominated by neutronic degrees of freedom and neutrons are well
bound with a separation energy of $\sim 8$ MeV, this may explain why a shell-model
calculation, within a restricted model space for a system close to the proton drip line,
gives a satisfactory agreement with the data.  
In order to reproduce the experimental 
$B(E2;4_1^+\rightarrow 2_1^+)$ transition, we obtained 
effective charges from our shell-model wave functions of 
$e_n=0.8-1.0e$ and $e_p=1.4e\pm 0.2e$. 
Our microscopically calculated effective charges are however slightly smaller,
$e_n=0.5-0.7e$ and $e_p=1.1-1.2e$.

\begin{table}[t]
\begin{center}
\caption{Proton and neutron effective charges relative to $^{100}$Sn for
the $1d_{5/2}$, $0g_{7/2}$, $1d_{3/2}$, $2s_{1/2}$ and $0h_{11/2}$
single particle orbitals.}
\begin{tabular}{lll}\\
& Proton & Neutron \\
\hline
$1d_{5/2}$ &1.06e &0.53e \\
$0g_{7/2}$ &1.15e &0.72e \\
$1d_{3/2}$ &1.04e &0.52e  \\
$0h_{11/2}$ &1.16e &0.51e \\
\end{tabular}
\label{tab:charges}
\end{center}
\end{table}

\end{multicols}

\end{document}